\documentclass[%
reprint,
showpacs,preprintnumbers,
amsmath,amssymb,
prl,
floatfix,
]{revtex4-1}

\usepackage{color}
\newcommand{\new}[1]{\textcolor{black}{#1}}

\usepackage{graphicx}
\usepackage{dcolumn}
\usepackage{bm}
\usepackage{hyperref}

\begin{document}
	
	
	\title{
		Footprints of hyperfine, spin-orbit, and decoherence effects in Pauli spin blockade}
	
	\author{T. Fujita$^{1}$}
	\author{P. Stano$^{2,3}$}
	\author{G. Allison$^{1,2}$}
	\author{K. Morimoto$^{1}$}
	\author{Y. Sato$^{1}$}
	\author{M. Larsson$^{1}$}
	\author{J.-H. Park$^{2}$}
	\author{A. Ludwig$^{4}$}
	\author{A. D. Wieck$^{4}$}
	\author{A. Oiwa$^{5}$}
	\author{S. Tarucha$^{1,2}$}
	\affiliation{$^1$Department of Applied Physics, The University of Tokyo, 7-3-1 Hongo, Bunkyo-ku, Tokyo 113-8656, Japan}
	\affiliation{$^2$Center for Emergent Matter Science (CEMS), RIKEN, 2-1 Hirosawa, Wako-shi, Saitama 351-0198, Japan}
	\affiliation{$^3$Institute of Physics, Slovak Academy of Sciences, 845 11 Bratislava, Slovakia}
	\affiliation{$^4$Lehrstuhl f\"{u}r Angewandte Festk\"{o}rperphysik, Ruhr-Universit\"{a}t Bochum, Universit\"{a}tsstra\ss e 150, Geb\"{a}ude NB, D-44780 Bochum, Germany}
	\affiliation{$^5$The Institute of Scientific and Industrial Research, Osaka University, 8-1 Mihogaoka, Ibaraki, Osaka 567-0047, Japan}
	
	
	\date{\today}
	
	\begin{abstract}
		We detect in real time inter-dot tunneling events in a weakly coupled two electron double quantum dot in GaAs. At finite magnetic fields, we observe two characteristic tunneling times, $T_d$ and $T_b$, belonging to, respectively, a direct and a blocked (spin-flip-assisted) tunneling. The latter corresponds to lifting of a Pauli spin blockade and the tunneling times ratio $\eta=T_b/T_d$ characterizes the blockade efficiency. We find pronounced changes in the \new{behavior of $\eta$ upon increasing the magnetic field, with $\eta$ increasing, saturating and increasing again.} We explain this behavior as due to the crossover of the dominant blockade lifting mechanism from the hyperfine to spin-orbit interactions and due to a change in the contribution of the charge decoherence. 
	\end{abstract}
	
	
	
	\maketitle
	
	
	Electron spins in semiconductor quantum dots are promising resources for quantum information processing \cite{loss1998:PRA,kloeffel2013:ARCMP}. Laterally gated dots \cite{hanson2007:RMP} are especially attractive due to the flexibility and scalability \cite{delbecq2014:APL} of their design, and the possibility to electrically initialize \cite{elzerman2004:N}, manipulate \cite{landriere2008:NP,yoneda2014:PRL}, and measure \cite{hanson2005:PRL,nowack2011:S} the \new{slowly relaxing} \cite{amasha2008:PRL,bluhm2011:NP} spin states. \new{Pauli spin blockade} (PSB) \cite{ono2002:S} plays a crucial role in electrical manipulations. PSB is established when the conservation of spin blocks a transition from an excited state, where two electrons in two dots have parallel spins, to the ground state, where they form a singlet in one dot. The spin can thus be detected by a local charge sensor as the presence or absence of a charge transition \cite{reilly2007:APL,cassidy2007:APL}. The blockade is lifted by \new{spin flips}, limiting the readout fidelities \cite{barthel2009:PRL,barthel2010:PRB}, as well as manipulations and preparations of quantum states.\cite{shulman2012:S,nichol2015:NC}
	
	In GaAs quantum dots, there are two important sources of \new{electron} spin flips: the spin-orbit coupling, and the hyperfine interaction with spins of atomic nuclei. Respectively, they dominate the spin relaxation time $T_1$\cite{khaetskii2001:PRB,johnson2005:N} and decoherence time $T_2$ \cite{khaetskii2002:PRL,petta2005:S,delbecq2016:PRL}. Apart from \new{causing detrimental effects}, both of these can be utilized in quantum state manipulation as a means of coupling of the electrical control fields to spins \cite{nowack2007:S,srinivasa2013:PRL,foletti2009:NP,wieck1984:PRL}. It is known that the relative importance of these two effects changes with magnetic field strength and orientation \cite{nadj-perge2010:PRB,raith2012:PRL,nichol2015:NC}.
	By resolving the direct and spin-flip-assisted inter-dot tunneling, here we investigate experimentally the limit these factors impose on the effectiveness of PSB. We find a crossover in their dominance upon changing the magnetic field strength, which is fully consistent with our theory. Our results give guidance on how to increase the PSB effectiveness with importance for spin readout applications.
	
	\begin{figure}
		\includegraphics[width=0.45\textwidth]{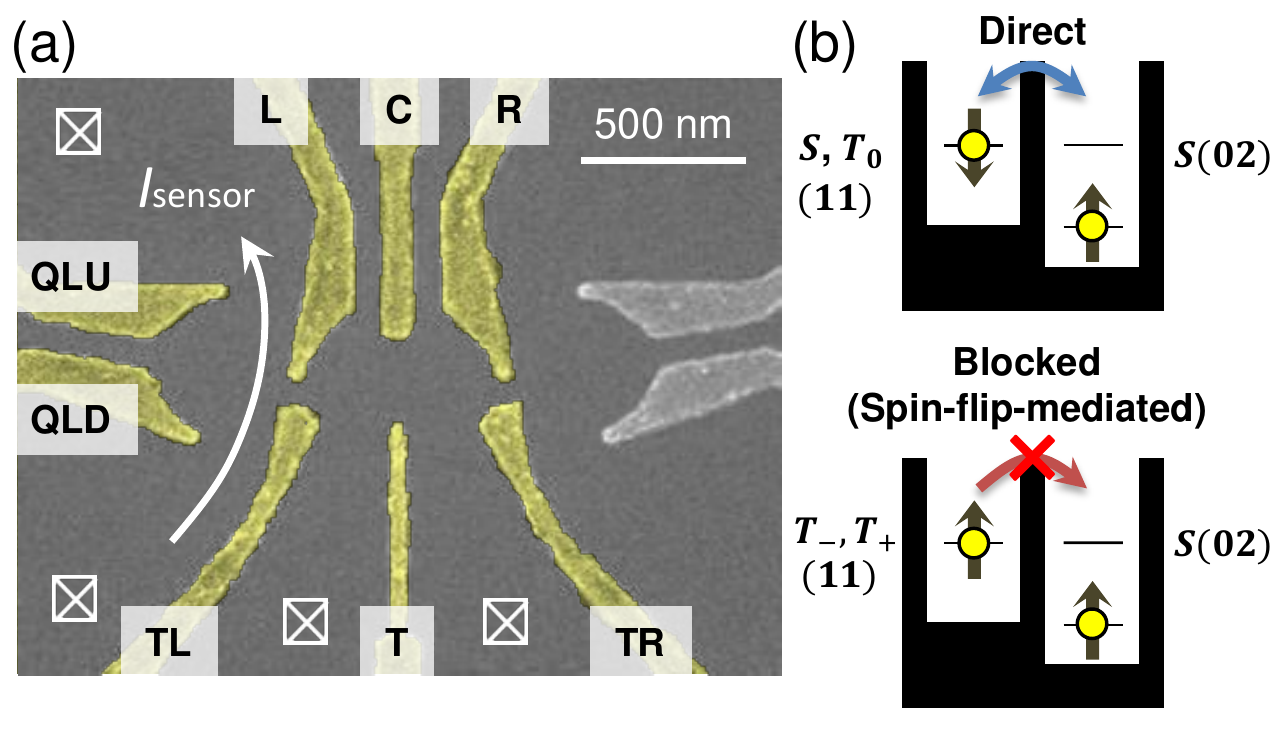}
		\caption{(color online). (a) A scanning electron micrograph of a sample similar to that measured.
			(b) Schematics of the direct and spin-flip-assisted inter-dot tunneling. \label{Fig_1} }
	\end{figure}
	
	Our device is a gate defined lateral double quantum dot (DQD), Fig.~\ref{Fig_1}(a), weakly tunnel-coupled and isolated from reservoirs, with lead-dot tunneling rates of order Hz, and the inter-dot tunneling rate of order kHz. In this regime, where tunnel coupling energies are much smaller than orbital or charging energies, the two electron configurations span a basis of five states \cite{coish2005:PRB,stepanenko2012:PRB}:
	one (02) charge state, the singlet $S(02)$, and four (11) charge states, the two spin polarized triplets $T_\pm(11)$, the unpolarized triplet $T_0(11)$, and the singlet $S(11)$. Here by $(N_L N_R)$ we denote the left and right dot occupancies as $N_L$ and $N_R$, respectively. Since the exchange energy splitting among the (11) charge states is negligible, the four (11) states are degenerate and, in general, energetically separated from the $S(02)$ state by the detuning energy $\Delta$. The nearby charge sensor can discriminate different charge states \cite{fujita2013:PRL}.
	Using gates L and R, we tune the dot close to the (11)-(02) degeneracy, $\Delta \approx 0$, by balancing the time averaged occupations of the two charge configurations, and measure the sensor current $I_{\rm sensor}$. With the inter-dot tunneling time set above the time resolution of the sensor, we monitor in this way the dot charge configuration in real time.
	
	\begin{figure}
		\includegraphics[width=0.45\textwidth]{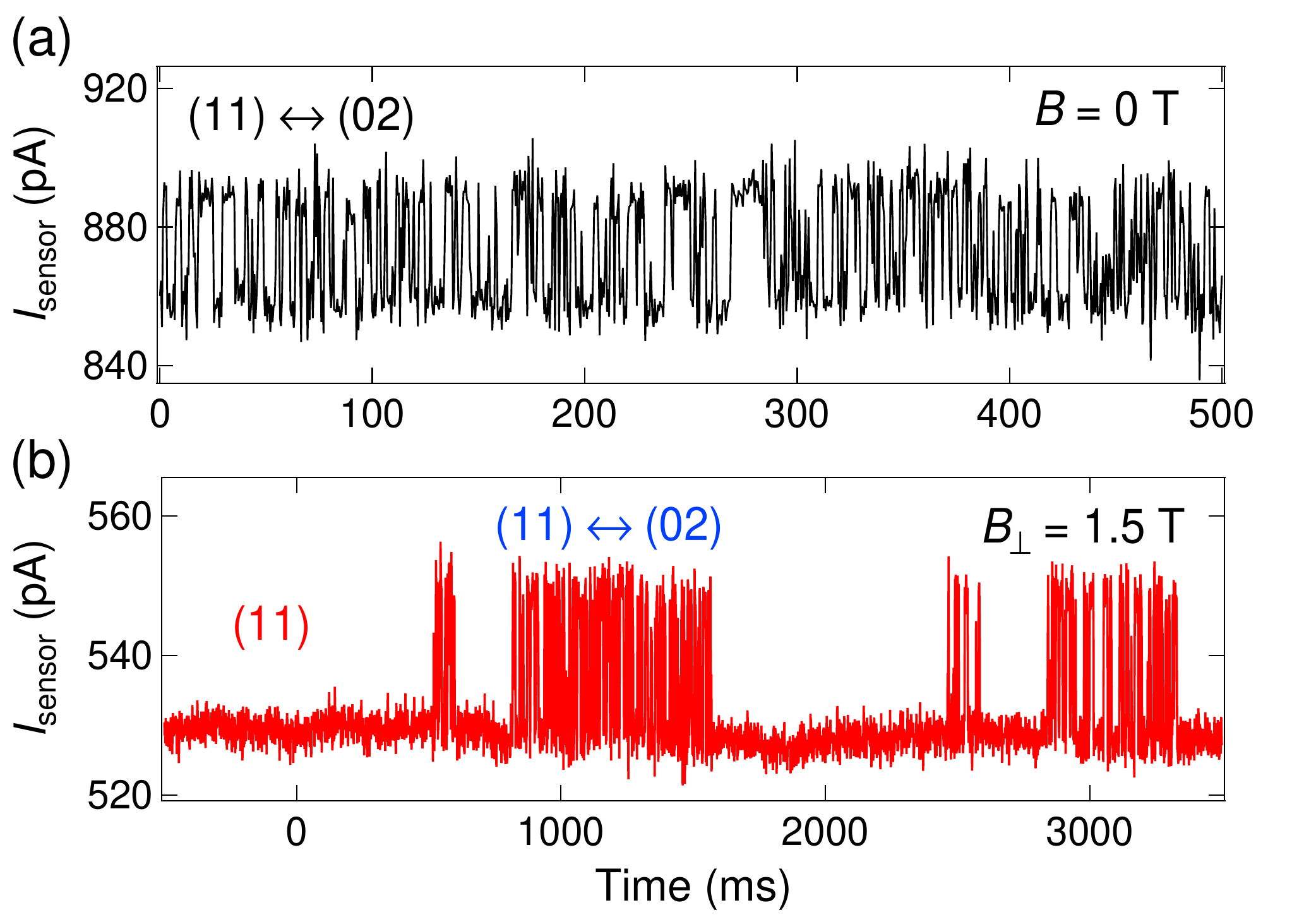}
		\caption{(color online). Typical real time charge sensor signals of a weakly tunnel coupled DQD at the (11)-(02) degeneracy for zero (a) and finite (b) magnetic fields. In (a) the signal shows repeated direct tunneling events, switching the dot between totally mixed spin states in (11) and the S(02) state. In (b), spin polarized (11) states are energetically split by a finite field and blocked in the (11) state, having a longer time to tunnel into the S(02) state.\label{Fig_2}}
	\end{figure}
	
	Figure 2(a) shows random (thermally excited) switching of charge configurations \cite{Pazy2004NanotechnologyIEEETransactionson} at zero magnetic field. The histogram of $(11)$ to $(02)$ tunneling times plotted on the upper left panel of Fig.~3 shows that the tunneling is described by a single time constant $T$, with the probability that no tunneling occurs for time $\delta t$ being $\exp(-\delta t/T)$. 
	Despite different spin configurations of the (11) states, a single tunneling rate into the (02) singlet is expected due to the hyperfine field of nuclear spins. Indeed, if described as a Zeeman term of a slowly fluctuating classical magnetic field located in the left(right) dot ${\bf B}_N^{L(R)}$ \cite{neder2011:PRB}, these quasi-static random fields in general couple all five states. Though the couplings between the (11) and (02) states are negligible, 
	they are appreciable among the (11) states (see below). As a consequence, no matter in which (11) spin state the system starts, within a few nanoseconds it contains the $S(11)$ state with \new{an} amplitude of order 1 from where it can tunnel to $S(02)$. As a result, within our time resolution, all (11) states tunnel out with the same rate and there is no PSB.
	
	The charge switching behavior is different at a large enough external magnetic field ${\bf B}$, see Fig.~2(b) for $B_\perp=1.5$ T. In addition to the fast switching \new{as in} Fig.~2(a), there are long intervals where the system remains in a (11) state. This is the Pauli spin blockade: once $B>B_N$, the Zeeman energy offset of the polarized triplets suppresses their hyperfine induced admixture with $S(11)$ and by that their tunneling to $S(02)$. Since $T_0(11)$ still mixes fast with $S(11)$, we expect to see two tunneling times, $T_b$ for spin-flip-assisted tunneling of spin polarized states, and  $T_d$ for direct tunneling of spin unpolarized states. The two processes are sketched on Fig~\ref{Fig_1}(b), and the histograms plotted in Fig.~3 indeed show bi-exponential distributions for $B_{||} \geq 100 $ mT. 
	
	\begin{figure}
		\includegraphics[width=0.45\textwidth]{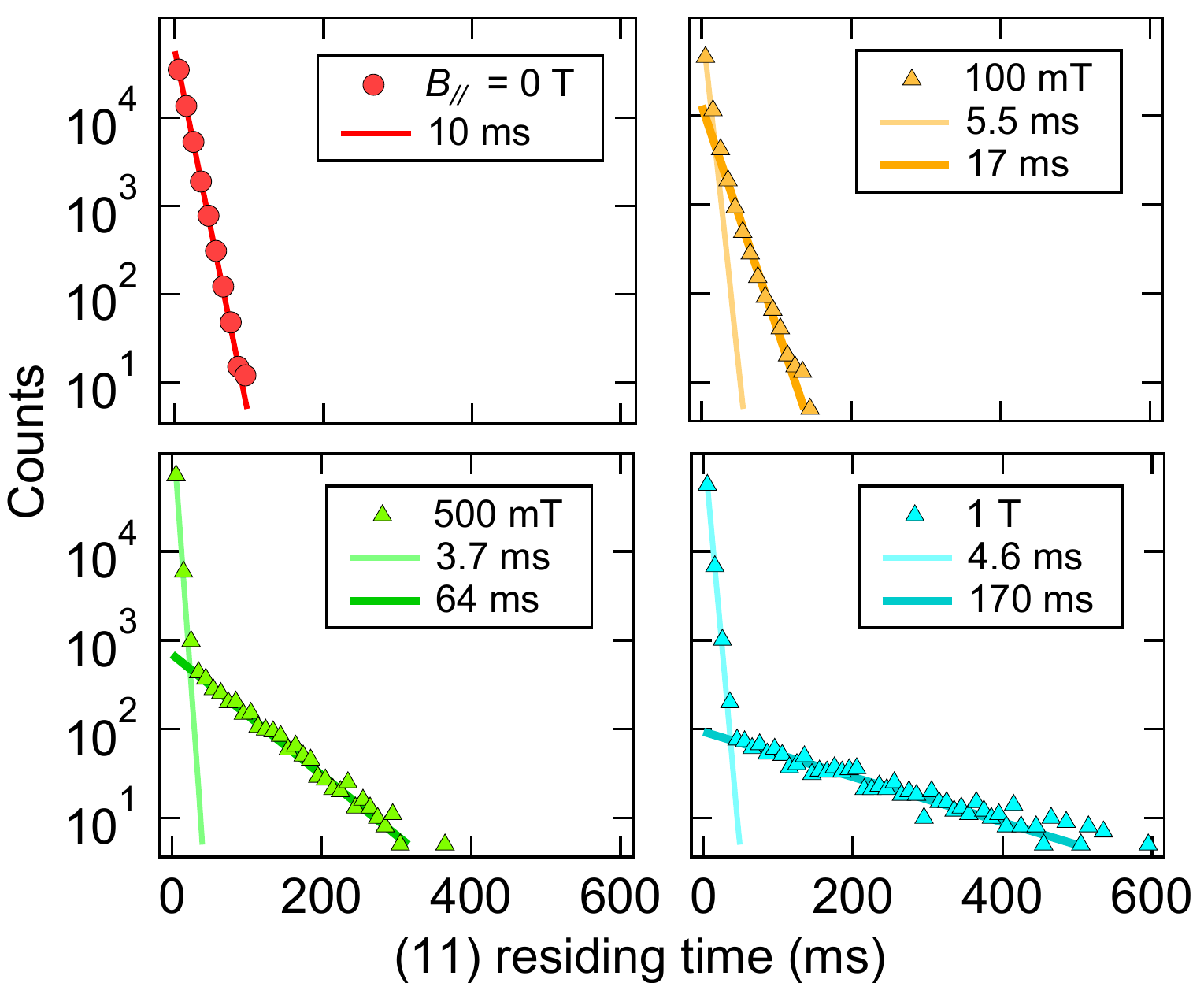}
		\caption{(color online). Example histograms of the (11) charge state residing time for different in-plane magnetic fields. Lines show the fitted linear trend for zero field and at short, $T_d$, and long, $T_b$, times for finite fields (see SM for details of the fitting procedure).
			\label{Fig_3}}
	\end{figure}
	
\begin{figure}
		\includegraphics[width=0.45\textwidth]{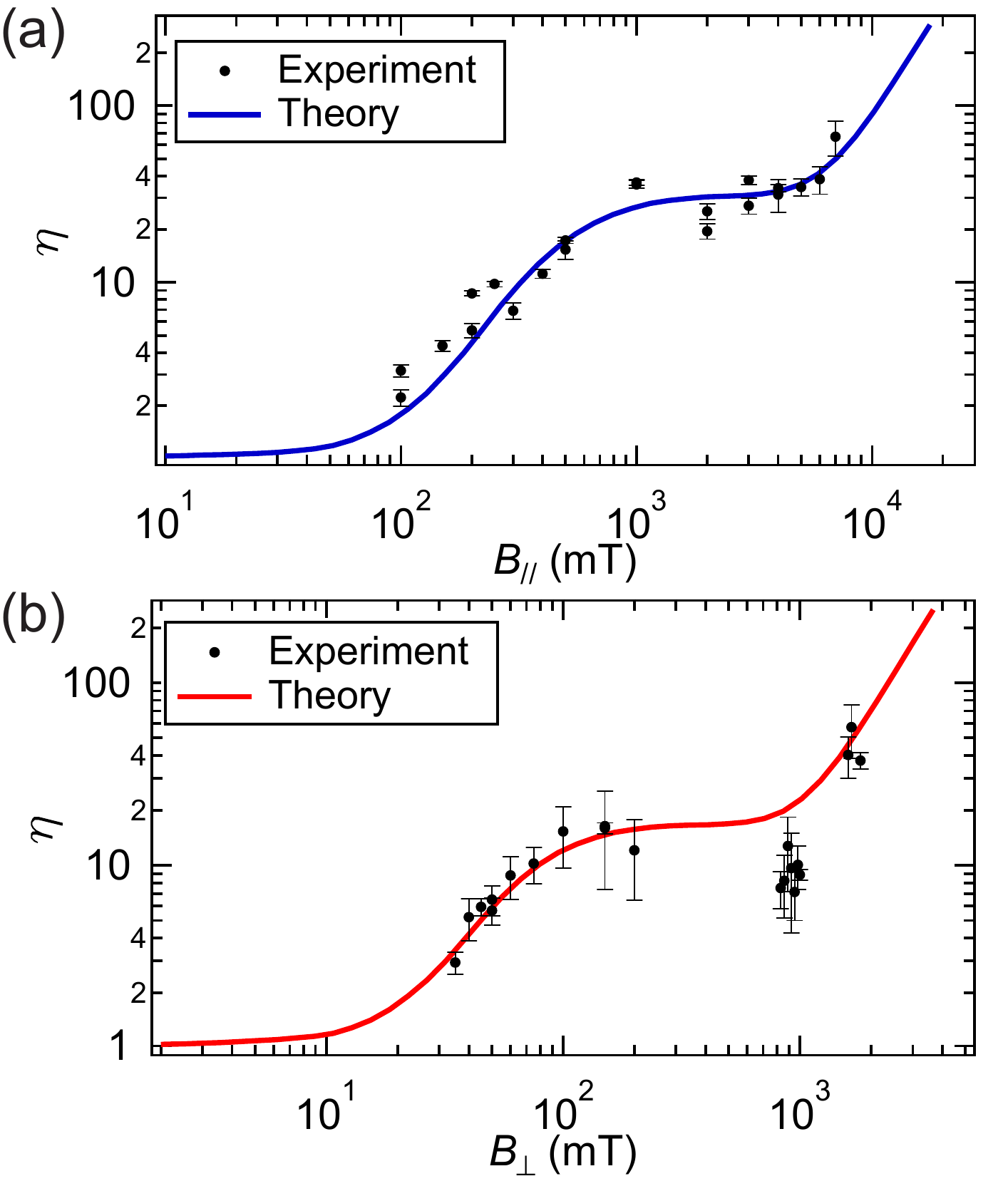}
		\caption{(color online). The ratio of tunneling times \new{$\eta$} as a function of, respectively, parallel (a), and perpendicular (b), magnetic field. Lines are fits using the model and parameters described in the text.
			\label{Fig_4}}
	\end{figure}

	We have investigated this PSB manifestation as a function of the magnetic field strength. In search for generic features, we measured for various sample cooldowns, which can change the shape of the dot, and for both in-plane and perpendicular magnetic fields, by which we isolate the strong orbital effects of the latter. Due to the influence of the AlGaAs barriers \cite{allison2014:PRB}, the $g$-factor is small, $|g_{\perp}| < 0.12$ for an out-of-plane magnetic field, and, as we find from the analysis below, about five times smaller for in-plane fields. (These small values make it easier to analyze the behavior of the rates at small Zeeman energies, which are pushed to higher magnetic fields by the small $g$-factors.) Because \new{of variations in the} measurement conditions, there is little systematic dependence of the tunneling times taken individually (see Fig.~S2 in Ref.~\cite{SM}). This is mostly due to the exponential sensitivity of the $S(11)-S(02)$ tunneling matrix element $\tau$, which is hard to keep constant during the re-alignement of the states' energies required in the measurement course. However, in plotting the ratio $\eta=T_b/T_d$ as in Fig.~\ref{Fig_4}, $\tau$ drops out and a clear trend emerges. 
	
	From zero to moderate external fields, $\eta$ behaves as expected: initially equal to one, it offsets once the external field becomes larger than the nuclear field. Here it grows as $\eta\propto (B/B_N)^2$, as is well known \cite{johnson2005:N} and can be understood from a simple perturbation theory (see below). At higher fields, however, we find that the growth stops and $\eta$ saturates. This is not completely unnatural, as it suggests that the PSB  effectiveness is limited by some process, expected to be eventually the case. However, moving to even higher fields, $\eta$ increases again \footnote{We can reach the last upturn in the trend unequivocally only for out-of-plane fields, where the $g$-factor, and the corresponding Zeeman energy, is five times larger than for the in-plane field.}. This is, however, completely surprising, as it implies that the limitation disappears. It is also at odds with the general behavior of the spin (inelastic) relaxation time $T_1$ between Zeeman split states, which is known to decrease with the magnetic field (as $B^{-5}$) as was predicted in theory and confirmed experimentally \cite{khaetskii2001:PRB,amasha2008:PRL}.

	To understand this peculiar behavior, it is necessary to consider several ingredients. To this end, we consider a Hamiltonian comprising the part defining the double dot $H_0$, and the interactions with the external, hyperfine, and spin-orbit fields \cite{jouravlev2006:PRL,danon2010:PRB}:
	\begin{equation}
		H=H_0 + H_Z + H_{nuc} + H_{so}.
		\label{eq:H}
	\end{equation}
	With details in Ref.~\onlinecite{SM}, $H_0$ consists of the electron kinetic energy, confinement potential, and Coulomb interaction and defines the Hilbert space as described above, with the (11) states detuned from $S(02)$ by $\Delta$ and the two singlets tunnel coupled by $\tau\equiv \langle S(11) | H_0 | S(02) \rangle$. 
	The spin-polarized triplets are offset by the Zeeman energy $\pm |g \mu_B B|$, which for our $g$-factors corresponds to 7 $\mu$eV for $B_\perp=1$ T and  0.7 $\mu$eV for $B_{||}=1$ T. A typical matrix element of $H_{nuc}$ within the (11) subspace is of order 0.1 $\mu$eV. Finally, assuming $H_{so}$ contains the linear-in-momentum Dresselhaus and Rashba terms, the only non-zero matrix elements are
	\begin{equation}
		\langle T_\pm(11) | H_{so} | S(11) \rangle = \pm \sqrt{2} g \mu_B B \frac{d}{\lambda_{so}}.
		\label{eq:soime}
	\end{equation}
	Here, $\lambda_{so}$ is an effective spin-orbit length, a combination of the Dresselhaus and Rashba coefficients, and $d$ is half of the interdot distance, which we estimate to be 130 nm from typical values of $T_d$ \cite{SM}.
	
	The system dynamics is given by the equation for its density matrix $\rho$,
	\begin{equation}
		i \hbar \partial_t \rho = [H, \rho] + L [\rho].
		\label{eq:rho}
	\end{equation}
	The last term is due to charge noise, usually dominated by fluctuating electric fields of phonons, gate potentials, impurities, and the charge sensor current. It leads to a fast decay of charge superpositions, with the decoherence rate $\Gamma$, typically several GHz \cite{petersson2010:PRL}. Denoting a (11) state as $X$, and $S(02)$ as $S$, the charge decoherence is described by $(L[\rho])_{XS} = -\Gamma \rho_{XS}$, a form independent of the (11) subspace basis. We use this invariance to simplify Eq.~\eqref{eq:rho} by choosing basis states $|X\rangle$ in which $H$ is diagonal within the (11) subspace. Because of the hyperfine and spin-orbit couplings, these eigenstates are in general superpositions of all four (11) states. In this basis, the remaining off-diagonal matrix elements are the tunneling terms
	\begin{equation}
		\tau_{XS} \equiv \langle X | H | S(02)\rangle =  \tau \langle X| S(11)\rangle,
		\label{eq:hxs}
	\end{equation}
	which are much smaller than the states' energy differences and can be treated perturbatively. In the leading order, the dynamics given by Eq.~\eqref{eq:rho} reduces to transitions between $X$ and $S$ with the rate
	\begin{equation}
		T_X^{-1} = \frac{2\pi}{\hbar^2} \int {\rm d}\omega \frac{1}{\pi} \frac{\Gamma}{\Gamma^2 + (\omega-\omega_{XS})^2} \, H_{XS}^2(\omega).
		\label{eq:tau}
	\end{equation}
	For further convenience, we introduce the spectral density of the transition matrix element
	\begin{equation}
		H_{XS}^2(\omega) =\frac{1}{2\pi} \int {\rm d}t \langle X | H(0) | S\rangle \langle S | H(t) | X \rangle \exp (i\omega t).
		\label{eq:tme}
	\end{equation}
	For a time independent Hamiltonian, which we consider at the moment, $H_{XS}^2(\omega)=|\tau_{XS}|^2\delta(\omega)$. Inserting this into Eq.~\eqref{eq:tau} gives Fermi's Golden rule formula with the initial and final state difference $E_X-E_S=\hbar \omega_{XS}$, Lorentzian\new{-}broadened by the decoherence. 
	
	We fit this model to the data on Fig.~\ref{Fig_4} by averaging the rates given by Eq.~\eqref{eq:tau} over hyperfine fields ${\bf B}_N^{L,R}$, assuming the latter having a Gaussian probability distribution with zero mean value and dispersion $B_N^2$. The typical hyperfine field $B_N$, the charge decoherence rate $\Gamma$, and the spin-orbit length $\lambda_{so}$ are fitting parameters. To take into account the limited accuracy of the detuning and the voltage jitter present in real experiments, we average over $\Delta \in \langle 0, 12 \rangle\, \mu$eV, a range corresponding to the electron temperature \footnote{Our resolution is limited due to the finite electron temperature (100 mK), gate potential resolution (1 $\mu$eV), and device stability. A small offset $\Delta=k_B T \ln 4$ is expected because of the fourfold difference in the spin degeneracy of the (11) and (20) charge states.}. However, the influence of $\Delta$ on the data fit in Fig.~4 is relatively minor.
	On the contrary, $B_N$, $\Gamma$, and $\lambda_{so}$ have a profound influence and the energy
	scales connected to these key quantum dot spin qubit parameters can be directly read off from the magnetic field dependence of $\eta$, as we now explain.
	
	Let us first assume the detuning is zero, and the hyperfine fields are fixed. Using Eqs.~\eqref{eq:hxs}--\eqref{eq:tme} we get the ratio of tunneling times of two (11) states $X(11) \leftrightarrow S(02)$, $Y(11) \leftrightarrow S(02)$ as
	\begin{equation}
		\frac{T_X}{T_Y} = \frac{\Gamma^2+\omega_{XS}^2}{\Gamma^2+\omega_{YS}^2} \left| \frac{\langle Y| S(11)\rangle}{\langle X| S(11)\rangle} \right|^2.
		\label{eq:explanation}
	\end{equation}
	Take first $B=0$. As already explained, the hyperfine fields fully mix the (11) subspace, so that each eigenstate typically contains the same amount of admixture of $S(11)$. Because in addition the Zeeman energy of hyperfine fields is negligible compared to $\hbar \Gamma$, the ratio in Eq.~\eqref{eq:explanation} is 1. Once $B \gg B_N$, the singlet admixture into the Zeeman split triplets is small, $|\langle X | S(11)\rangle |^2 \propto B_N^2/B^2$. The remaining two states, typically equally mixed $S(11)$ and $T_0(11)$, have $|\langle Y | S(11)\rangle |^2 \sim 1/2$. This gives two tunneling times with the ratio proportional to $B^2/B_N^2$.
	
	The \new{plateau terminating the growth} of $\eta$ at higher fields (around 0.1 T for $B_\perp$) can be understood as the spin-orbit taking over the hyperfine field in the matrix element in Eq.~\eqref{eq:hxs}. Indeed, whereas the latter is independent of $B$, the former grows linearly, see Eq.~\eqref{eq:soime}. This equation also gives the spin-orbit length as $\lambda_{so} \sim 2d \sqrt{2\eta}$ with $\eta$ the ratio on the plateau.
	
	Increasing the magnetic field further (beyond 1 T for $B_\perp$), $\eta$ starts to grow again in Fig.~4. This can be still reconciled with Eq.~\eqref{eq:explanation}, as due to the first fraction on its right hand side. Namely, once the Zeeman energy becomes larger than the decoherence, the spectral overlap of spin-polarized $(11)$ states and $S(02)$ drops compared to spin-unpolarized $(11)$ states. The Zeeman energy where $\eta$ starts increasing for the second time gives therefore the charge decoherence rate $\hbar \Gamma$.
	
	The three energy scales extracted visually as just described from the slope changes of $\eta$ give the values of parameters $B_N$, $\lambda_{so}$, and $\Gamma$ within a factor of order one. 
	We found that the best way to nail down these factors quantitatively is straightforward numerics. Namely, for given values of hyperfine fields, we diagonalize the $4\times4$ Hamiltonian in the (11) subspace numerically, and calculate the rates according to Eq.~\eqref{eq:tau}. We average these over typically 10$^6$ hyperfine fields random configurations. Because we cannot distinguish experimentally all four rates, we define in our numerics the ``blocked(direct)'' rate as the average of the first(last) two rates ordered by their magnitudes. 
	
	In this way, we obtain the solid lines in Fig.~\ref{Fig_4} using $|g|\mu B_N=1.7\,\mu$eV, $\Gamma=7$ GHz, and $\lambda_{so}=1.1\,\mu$m for the out-of-plane field, and $\lambda_{so}=1.5\,\mu$m for the in-plane field. From the value of $B_N$, we can infer the number of nuclei within the dot volume \cite{merkulov2002:PRB}, $N=(A I(I+1)/g\mu_B B)^2\approx 1.2\times 10^5$, using $A=90$ $\mu$eV, and $I=3/2$. All extracted values are typical for gated dots in GaAs, that is $N$ \cite{taylor2007:PRB}, the charge decoherence rate \cite{hayashi2003:PRL,petta2004:PRL}, and spin-orbit lengths \cite{zumbuhl2002:PRL,amasha2008:PRL}. We note that the different values of the effective spin-orbit length fitted for in-plane and out-of-plane magnetic fields are consistent with directional anisotropies of $\lambda_{so}$ \cite{stano2006:PRL}, observed in dot spectra \cite{nichol2015:NC}, and spin relaxation \cite{scarlino2014:PRL}.

	We also considered alternative explanations, examining inelastic (11) to (02) transitions due to a non-dipolar electric noise, inelastic ($T_1$) transitions within the (11) subspace, and lifting the spin-blockade by cotunneling. As none of these can be naturally reconciled with the data, we give these details only in Ref.~\cite{SM}.
	
	We conclude by suggesting how to increase the PSB effectiveness.
	The spin-orbit effects should be minimized, what can be achieved by orienting the magnetic field along certain in-plane directions \cite{malkoc2016:PRB}, specified by setting $\lambda_{so}^{-1}=0$ in Eq.~(S23) in Ref.~\cite{SM}. We predict that the quadratic growth $\eta \sim B^2$ will then extend to much higher fields and increase to $B^4$ once the Zeeman energy becomes larger than the charge decoherence rate. Finally, these properties are to a large extent independent of the value of the interdot tunneling,  \new{increase of} which should therefore boost both direct and blocked rates while preserving their ratio.
	
	This work was supported by Grant-in-Aid for Scientific Research 
	S (26220710) and A (25246005), Innovative area (26103004), CREST, JST, ImPACT program, 
	the IARPA project  ``Multi-Qubit Coherent Operations'' through Copenhagen University, MEXT Project for Developing Innovation Systems and QPEC, The University of Tokyo. T. F. \& H. K. are supported by JSPS Research Fellowships for Young Scientists. P.~S. acknowledges the support of APVV-0808-12(QIMABOS). A.L. and A.D.W. acknowledge gratefully support of DFG-TRR160 and BMBF-Q.com-H 16KIS0109.
	
%

\newpage

\renewcommand{\theequation}{S\arabic{equation}}
\renewcommand{\thefigure}{S\arabic{figure}}
\renewcommand{\bibnumfmt}[1]{[S#1]}
\renewcommand{\citenumfont}[1]{S#1}

	\newpage
	
	\appendix
	
	\begin{center}
		{\Large \bf Supplemental Information}
	\end{center}
	
	Here we give details on device characteristics, additional measurements, and the derivation of formulae. In Sec.~I we give additional information on the device, comment on the $g$-factor anisotropy, and spontaneous (inelastic) spin flip by phonon emission. In Sec.~II we show measurement data demonstrating Pauli spin blockade in a configuration where there is a voltage bias across the dot. In Sec.~III we describe the fitting procedure used to extract the two tunneling times $T_b$ and $T_d$, resulting in the bi-exponential fits shown in Fig.~3 of the main text. The rest contains details on the theory. In Sec.~IV we give the explicit form of the two electron double dot Hamiltonian, the restricted Hilbert space, and the corresponding matrix elements of the hyperfine and spin-orbit interactions. In Sec.~V we derive Eq.~(5). In Sec.~VI we comment on alternative explanations of the data we considered.
	
	\section{I.~Additional information on the device}
	
	\subsection{Experimental details}
	Single electrons are captured by confining a two-dimensional electron gas (2DEG) formed in an AlGaAs/GaAs/AlGaAs double heterostructure quantum well with a GaAs width of 7.3 nm for small electron $g$-factor. The 2DEG has a carrier density of 2.1 $\times$ 10$^{11}$ cm$^{-2}$, and a mobility of 1.0 $\times$ 10$^5$ cm$^2$/Vs. The upper (lower) barrier is 95 nm (100 nm) Al$_{0.34}$Ga$_{0.66}$As. The top 65 nm in the upper barrier is doped by Si. The device is placed in a dilution refrigerator at a base temperature of 25 mK. The double dot charge configuration is monitored by measuring a charge sensor current using a room temperature current amplifier with a rise-time of 100 $\mu$s.

	\subsection{ $g$-factor anisotropy}
	In Fig.~4 of the main text, the dependence of $\eta=T_b/T_d$ on the magnetic field magnitude $B$ shows three crossovers in the trend. The values of $B$ for these crossovers differ for in-plane and perpendicular orientation of the field due to the $g$-factor anisotropy, since the Zeeman energy, which determines the energy scale of the crossover fields, is proportional to the $g$-factor. The g-factor is reduced compared to that of the bulk due to the small thickness of the GaAs layer and Al content of the barriers in our heterostructure  \cite{Jeune1997SemiconductorScienceandTechnology}. 
	
	The g-factor of the 2DEG used in this work, $\left|g_\perp\right|=0.12$, was measured in a perpendicular magnetic field using an electron spin resonance technique \cite{allison2014:PRB}. The ratio $\left|g_\perp/g_{||}\right|=5$ extracted from the fitted data of Fig.~4 results in $\left|g_{||}\right|=0.024$, which is consistent with the $g$-factor anisotropy found in Ref.~\cite{Jeune1997SemiconductorScienceandTechnology} and with the fact that the $g$-factor of a 2DEG and a dot are comparable, concluded in Ref.~\cite{allison2014:PRB}.

	\subsection{Effect of spontaneous phonon emission (inelastic tunneling)}
	The electron spin relaxation rate due to spontaneous phonon emission is known to grow with the magnetic field magnitude. We do not see such an effect in our data, observing only an increase of $\eta$ with the field magnitude. Also, as we analyze in detail in Sec.~VI, such an inelastic transition within the (11) subspace [for example $T_-(11) \to S(11)$] is expected to result in a histogram of tunneling times qualitatively different from our observations (being non-linear at long times, see Fig.~S3, versus linear at long times as seen in Fig.~3 of the main text). 
	
	The absence of this effect is most probably due to the corresponding relaxation time being too long. This, in turn, is due to the small $g$-factor in our sample. Based on the previous works \cite{Amasha2008Phys.Rev.Lett.,stano2006:PRB}, we estimate the spin inelastic relaxation time for the highest fields we used to be of the order of seconds. This is much smaller than the longest inter-dot tunneling times that we observe (and model as elastic tunnelings), of the order of 100 ms. One expects that the effects of inelastic transitions would become appreciable in $\eta$ making the $g$-factor larger, and/or the inter-dot tunneling smaller.

	\section{II.~Pauli spin blockade in a charge diagram}
\begin{figure}
		\includegraphics[width=0.45\textwidth]{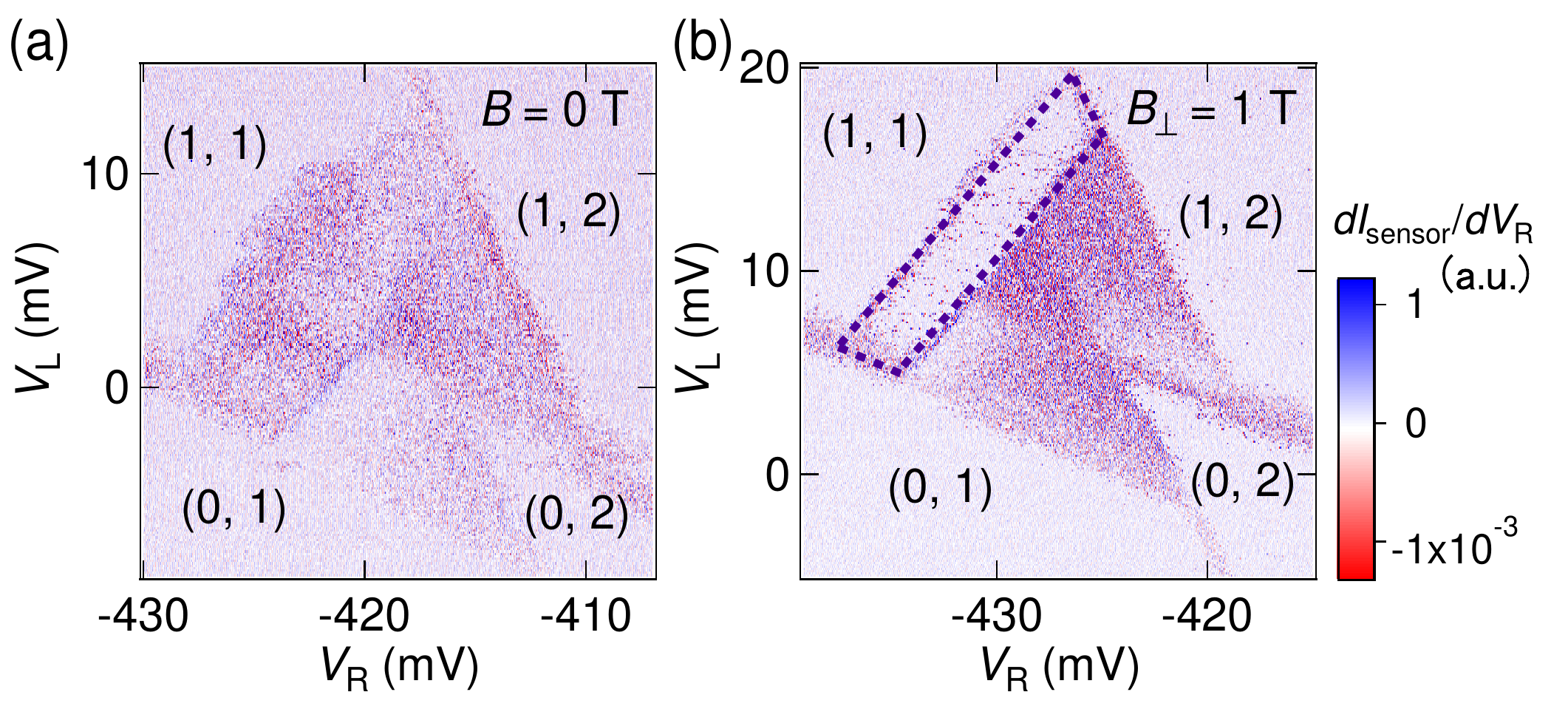}
		\caption{(color online). (a) Typical charge stability diagram around the (11)-(02) charge degeneracy in a dot biased by 1.4 mV with the dot-lead and inter-dot tunneling rates of the order of kHz. (b) Similar diagram at a finite magnetic field (here at $B_\perp$ = 1 T) where Pauli spin blockade is observed as a suppressed signal inside the highlighted trapezoidal region. \label{Fig_S1}}
	\end{figure}

	Here we show the charge diagram demonstrating the Pauli spin blockade in a weak tunnel coupling regime. To obtain it, we tuned both interdot and dot-lead tunneling rates to the order of kHz, applied a voltage of 1.4 mV across the dot, and swept through the displayed range in gate voltages $V_L$ and $V_R$. At zero magnetic field the charge stability diagram shows typical finite bias triangles, Fig.~\ref{Fig_S1}(a), where electrons sequentially tunnel through the DQD \cite{hanson2007:RMP}. The number of tunneling events per pixel is of the order of one, since the gate voltage sweep rate is comparable to the tunneling rates. Under such conditions, charge tunneling is observed as a noisy signal in the diagram. At finite magnetic field, a trapezoidal region of blocked current is developed at the base of the triangles, Fig.~\ref{Fig_S1}(b), a hallmark of Pauli spin blockade \cite{johnson2005b:PRB}. It reflects the trapping of the system in a polarized triplet state, where the electron cannot transit across the double dot without a spin flip, corresponding to blocked tunneling sketched on Fig.~1(b).
	
	\section{III.~Bi-exponential fitting}
	
	We extract the two (blocked and direct) tunneling times by fitting to a bi-exponential distribution in the following way. For each value of the magnetic field (one point in Fig.~4) we collect a set of $N$, typically $10^3$ to $10^4$, individual (11) to (02) tunneling times (such a set is plotted as a histogram in Fig.~3). We order the tunneling times in the set in ascending order, labeling them by integer index $i=1,\ldots,N$, so that $t_i<t_{i+1}$. We define the cumulative probability function $P_{data}(t_i)=i/N$, which describes the probability that no tunneling occurred until time $t$ as $1-P(t)$.
	In the model where the tunneling from (11) to (02) proceeds with a rate dependent on the (11) state (blocked or non-blocked, with initial occupation probabilities $p_b$ and $p_d$, respectively) we have
	\begin{equation}
		1-P_{model}(t)=p_b e^{-t/T_b}+p_d e^{-t/T_d},
	\end{equation}
	a sum of two exponential decays weighted by their probabilities. Assuming the system spends an equal amount of time in the blocked and unblocked configurations, we would have $p_b T_b=1/T_b+1/T_d$, and $p_d T_d=1/T_b+1/T_d$. However, to allow for possible imbalance and offset, we keep all four parameters $p_b,p_d,T_b$, and $T_d$ as fitting parameters. The fit means we find the values of these parameters which minimize the following sum
	\begin{equation}
		\sum_{i=1}^{N} [P_{model}(t_i)-P_{data}(t_i)]^2(t_i-t_{i-1})^n,
	\end{equation}
	where $n$ is typically set to 0.5$\sim$1.5. The value $n=1$ corresponds to the standard least square fitting of the discrete set of $P(t_i)$, whereas we found it useful to introduce different values of $n$ to weight tunneling times according to their density. This improves the fit at longer times where we have fewer tunneling events, without the necessity for fitting from a data histogram, which we found less robust to extract the slower tunneling rate. The fitting error of the extracted times $T_{b/d}$, defined by the standard confidence interval $95\%$, was below 1~\% therefore, to make the error bars visible on Fig.~4 we multiplied them by a factor of 50. This means that the error from the fitting is negligible compared to other errors (such as the potential drift due to the 1/f noise, the measurement errors due to noise in the charge sensor, etc.).

	The extracted individual tunneling times, $T_b$ and $T_d$, have less systematic dependence (Figs.~\ref{Fig_S2}(a) and \ref{Fig_S2}(b) taken for in- and out-of-plane $B$, respectively), compared to Fig.~4 where their ratios are plotted. The fluctuations of individual tunneling times are caused by fluctuations of the inter-dot tunnel coupling, in turn due to the influence of implementing changes in $B$, from deliberately changing gate voltages for other purposes in a related experiment,\cite{Fujita2013Phys.Rev.Lett.,Morimoto2014Phys.Rev.B}
	from the limited precision in setting the detuning $\Delta$ to zero, and---probably the most importantly---from the drift of local voltages due to the 1/f noise (it takes approximately 10 minutes to gather data for a single point in Fig.~4). Except for $\Delta$, the above mentioned noise effects do not appreciably influence the ratio $\eta$, since there the inter-dot tunnel coupling cancels out. The detuning is important if it, accidentally, matches the Zeeman energy, at which value the $S(02)$ anti-crosses with $T_+(11)$. This lifts the PSB of this spin-polarized state, and results in a relatively sharp dip in $\eta$. We took this effect into account by averaging over a distribution of $\Delta$, as described in the main text. A different probability distribution of $\Delta$, or its changes in the measurement course, might be the reason for the discrepancy with the fit, and also for the remaining fluctuations in $\eta$, which are much larger than the error bars obtained from the histogram fitting. Particularly, an accidental degeneracy of $S(02)$ and $T_+(11)$ might be the reason for the dip and large fluctuations of $\eta$ around $B_\perp \approx 1$ T.

	\section{IV.~Hamiltonian}
	
	\subsection{Explicit form}
	\begin{figure}
		\includegraphics[width=0.45\textwidth]{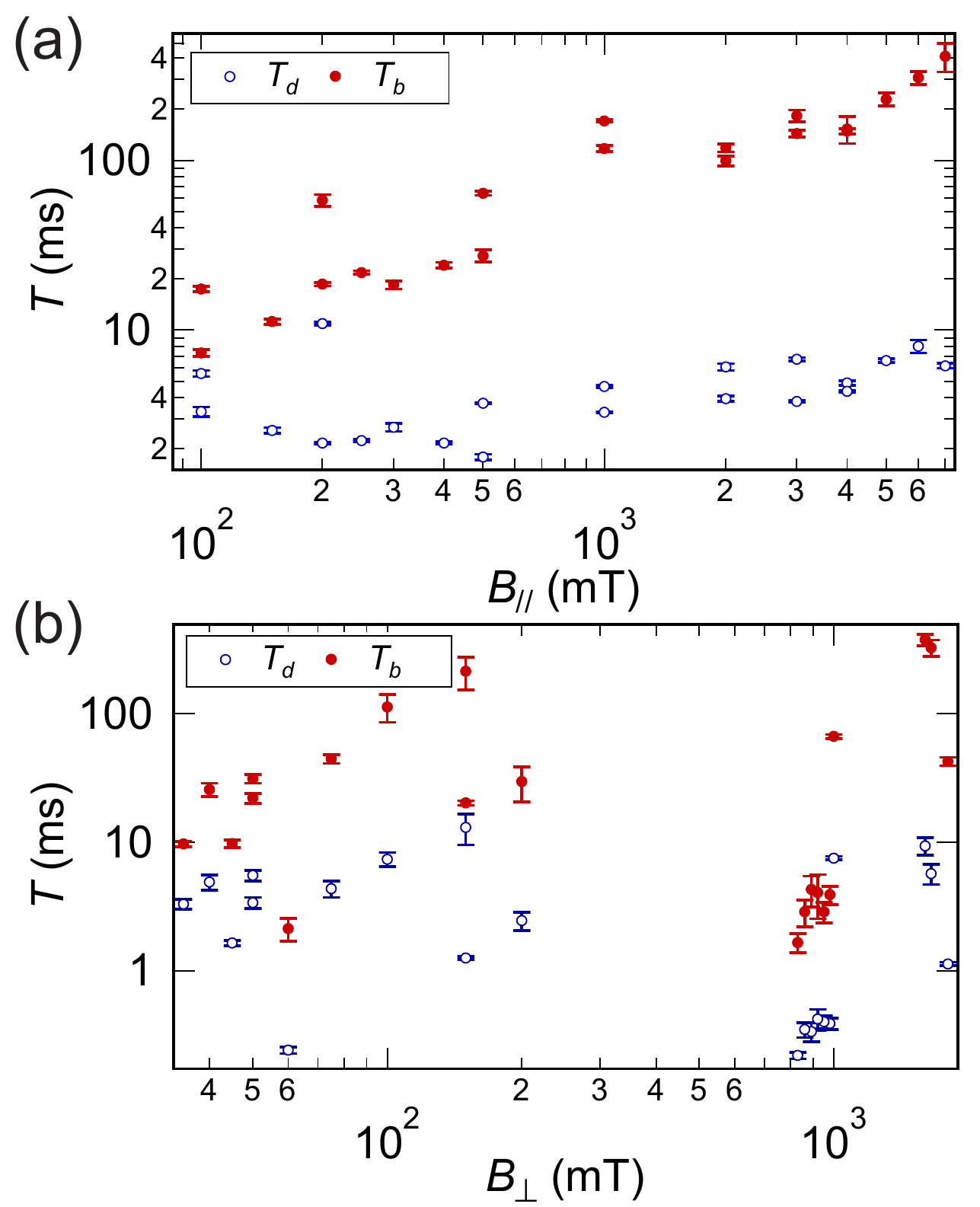}
		\caption{(color online). Individual tunneling times of direct, $T_d$, and spin-flip-assisted, $T_b$, tunneling. Values are extracted from bi-exponential curve fitting for (a) in-plane and (b) out-of-plane magnetic field. Taking the ratio $T_b/T_d$ results in data plotted in Fig.~4 of the main text. \label{Fig_S2}}
	\end{figure}
	
	Here we give the explicit form of the Hamiltonian of Eq.~(1). Denoting the $i$-th electron position within the 2DEG plane as ${\bf r}_i$, and its momentum operator as ${\bf p}_i=-i\hbar\boldsymbol{\nabla}_i+e {\bf A}$, the quantum dot Hamiltonian is
	\begin{equation}
		H_0 = \sum_{i=1,2} \left( \frac{{\bf p}_i^2}{2m} + V({\bf r}_i) \right) + \frac{e^2}{4\pi \epsilon_0 \epsilon_r} \frac{1}{|{\bf r}_1-{\bf r}_2|},
	\end{equation}
	with $m$ the effective mass, $e>0$ the elementary charge, $\epsilon_0$ the vacuum permittivity, $\epsilon_r$ the relative permittivity, and ${\bf A}=B_\perp(-y/2,x/2)$ the vector potential of the out-of-plane field. The confinement potential of a weakly coupled dot might be modeled, e.g., by the bi-quadratic form
	\begin{equation}
		V({\bf r}) = \frac{1}{2} m \omega ^2 {\rm min} \left\{ ({\bf r}-{\bf d})^2, ({\bf r}+{\bf d})^2 \right\},
		\label{eq:V}
	\end{equation}
	with $\hbar\omega$ the orbital energy scale, and the vector ${\bf d}=d (\cos \delta, \sin \delta)$ parametrizing the double dot geometry, with $2d$ the distance between the two potential minima (the inter-dot distance), and $\delta$ the angle between the dot main axis and the [100] crystallographic direction (close to either 45 or -45 degrees).
	
	The remaining terms of the total Hamiltonian, using the single electron notation, are the following. The Zeeman term is 
	\begin{equation}
		H_Z = \frac{g}{2} \mu_B {\bf B} \cdot \boldsymbol{\sigma},
		\label{eqS:Z}
	\end{equation}
	with $g$ the $g$-factor, $\mu_B$ the Bohr magneton, ${\bf B}$ the external magnetic field, and the electron spin operator $\boldsymbol{\sigma}$ is the vector of sigma matrices. The spin-orbit interaction is
	\begin{equation}
		H_{so} = \frac{\hbar}{m\lambda_D} \left( -p_x \sigma_x + p_y \sigma_y \right) + \frac{\hbar}{m\lambda_B} \left( p_y \sigma_x - p_x \sigma_y \right),
	\end{equation}
	with $\lambda_D$, and $\lambda_B$ the spin-orbit lengths corresponding to the Dresselhaus, and the Bychkov-Rashba interactions, respectively. Finally, the Fermi contact interaction is
	\begin{equation}
		H_{nuc} = \frac{A_0}{2} \sum_n v_0 |\psi_0(z_n)|^2 \delta({\bf r}-{\bf r}_n) \boldsymbol{\sigma} \cdot {\bf I}_n,
		\label{eqS:nuc}
	\end{equation}
	with $\psi_0$ the electron wave function in the perpendicular direction (the lowest subband of the 2DEG), and the index $n$ labels nuclear spins ${\bf I}_n$ at positions $({\bf r}_n,z_n)$, which have volume density $v_0^{-1}$. 
	
	The material parameters of GaAs are $m=0.067$ $m_e$, $\epsilon_r=12.9$, $\lambda_D$ and $\lambda_B$ typically of order $\mu$m, $A_0=90\,\mu$eV, $I=3/2$, and $v_0=a_0^3/8$, with the lattice constant $a_0=0.565$ nm. 
	
	\subsection{Low energy subspace of the two electron double dot Hamiltonian}
	
	For a weakly coupled double dot, the low energy subspace is well approximated using a basis built from the lowest Fock-Darwin solutions localized in the two potential minima,
	\begin{equation}
		\psi_{\mp d} ({\bf r})= \frac{1}{\sqrt{\pi} l_B} \exp\left( -\frac{({\bf r} \pm {\bf d})^2}{2l_B^2} \pm \frac{i e y B_\perp}{2\hbar}\right) ,
	\end{equation}
	with the magnetic field renormalized confinement length $l_B=[l_0^{-4}+B_\perp^2 e^2/4\hbar^2]^{-1/4}$. These can be orthonormalized to 
	\begin{equation}
		\Psi_{L/R} ({\bf r}) =  \psi_{\mp d} ({\bf r}) - \frac{\Omega}{2} \psi_{\pm d} ({\bf r}) + O(\Omega^2),
		\label{eqS:ono 1e basis}
	\end{equation}
	where we took the leading order in the left/right wave functions overlap,
	\begin{equation}
		\langle \psi_{L} | \psi_R \rangle = \Omega,
	\end{equation}
	which parametrizes the inter-dot coupling. For a weakly coupled dot, the overlap is much smaller than 1. 
	
	We now specify the two-electron states that constitute the basis of the restricted Hilbert space of the two electron Hamiltonian. Using the single electron orbitals $|L\rangle\equiv |\Psi_L\rangle$, and similarly for $R$, we construct the following five states
	\begin{subequations}
		\begin{eqnarray}
			|S(11)\rangle &=&\frac{1}{\sqrt{2}} | LR + RL \rangle \otimes \frac{1}{\sqrt{2}}  |\uparrow \downarrow - \downarrow\uparrow \rangle, \\
			|S(02)\rangle &=& | RR \rangle \otimes \frac{1}{\sqrt{2}}  |\uparrow \downarrow - \downarrow\uparrow \rangle,\\ 
			|T_0(11)\rangle &=&\frac{1}{\sqrt{2}} | LR - RL \rangle \otimes \frac{1}{\sqrt{2}}  |\uparrow \downarrow + \downarrow\uparrow \rangle, \\
			|T_+(11)\rangle &=&\frac{1}{\sqrt{2}} | LR - RL \rangle \otimes |\uparrow \uparrow \rangle, \\
			|T_-(11)\rangle &=&\frac{1}{\sqrt{2}} | LR - RL \rangle \otimes |\downarrow \downarrow \rangle,
		\end{eqnarray}
		\label{eqS:2e basis}
	\end{subequations}
	where we separated the orbital and spin parts of the wave function, for both of which we use the notation where the order of single particle wave functions denotes the state of the first and second electron, respectively. In the spin part, $|\uparrow\rangle$ and $|\downarrow\rangle$ denote a spinor oriented parallel and antiparallel to the external field ${\bf B}$. 
	
	\subsection{Matrix elements of the hyperfine fields}
	
	We can rewrite Eq.~\eqref{eqS:nuc} in the form of the Zeeman term, Eq.~\eqref{eqS:Z}, by introducing 
	\begin{equation}
		{\bf B}_{L/R} = \frac{2A_0}{g \mu_B} \sum_n  v_0 |\psi_0(z_n)|^2 |\Psi_{L/R}({\bf r}_n)|^2 {\bf I}_n,
	\end{equation}
	where the single electron wavefunctions $\Psi_{L/R}({\bf r}_n)$ are given in Eq.~\eqref{eqS:ono 1e basis}. Neglecting the small overlap $\Omega$, the two fields can be considered independent.
	
	Using this notation, $H_{nuc}$ enters through the average nuclear field ${\bf B}^n=({\bf B}_L+{\bf B}_R)/2$ as a contribution to the external magnetic field, generating the following matrix elements among the triplet states
	\begin{subequations}
		\begin{eqnarray}
			\langle T_\pm (11) | H_{nuc} | T_\pm (11) \rangle & = & \mp g \mu_B B^n_z,\\
			\langle T_\pm (11) | H_{nuc} | T_0 (11) \rangle & = & g \mu_B \frac{B^n_x \pm i B^n_y}{\sqrt{2}},
		\end{eqnarray}
		\label{eqS:Hnuc m.e. 1}
	\end{subequations}
	while the field gradient $\delta{\bf B}^n=({\bf B}_L-{\bf B}_R)/2$ generates matrix elements between the triplets and the singlet $S(11)$,
	\begin{subequations}
		\begin{eqnarray}
			\langle T_\pm(11) | H_{nuc} | S(11) \rangle & = & \mp g \mu_B\frac{\delta B^n_x \mp i \delta B^n_y}{\sqrt{2}},\\
			\langle T_0(11) | H_{nuc} | S(11) \rangle & = & g\mu_B\delta B^n_z.
		\end{eqnarray}
		\label{eqS:Hnuc m.e. 2}
	\end{subequations}
	The singlet $S(02)$ is coupled to other states by matrix elements which are, compared to those given, suppressed by a factor $\Omega$ and can be neglected.
	
	\subsection{Matrix elements of the spin-orbit interaction}
	
	To obtain the spin-orbit interaction matrix elements, we first unitarily transform the total Hamiltonian \cite{aleiner2001:PRL, rashba2003:PRB}  
	\begin{equation}
		H \to U^\dagger({\bf r}_1) U^\dagger({\bf r}_2) H U({\bf r}_1) U({\bf r}_2),
	\end{equation}
	with a unitary representing a position dependent spin rotation
	\begin{equation}
		U({\bf r}) = \exp\left( i {\bf n}_{so}({\bf r}) \cdot \boldsymbol{\sigma} \right),
	\end{equation}
	defined by the vector
	\begin{equation}
		{\bf n}_{so}({\bf r}) = \left( \frac{x}{\lambda_D} + \frac{y}{\lambda_B}, -\frac{x}{\lambda_B} - \frac{y}{\lambda_D}, 0 \right),
		\label{eqS:nso}
	\end{equation}
	expressed through the crystallographic coordinates $x$, $y$, and the spin-orbit lengths $\lambda$. In the leading order of the small quantity $d/\lambda$, the only effect of the transformation is to express the spin-orbit interaction in terms of the position, rather than momentum, operator. That means, the transformed Hamiltonian has the same form as the one in Eq.~(1), if the spin-orbit integration is replaced by
	\begin{equation}
		H_{so}^{eff}=g\mu_b \left( {\bf B} \times {\bf n}_{so} \right) \cdot \boldsymbol{\sigma}.
		\label{eqS:effective soi}
	\end{equation}
	This form is better suited for a perturbative treatment of the spin-orbit effects, as was demonstrated in Refs.~\cite{baruffa2010:PRL, baruffa2010:PRB}. We note that the transformation leaves the statistics of the random nuclear fields unchanged to all orders of $d/\lambda$, as a local rotation of a quantization axis is irrelevant for unpolarized spins (only the correlators of distant spins are different in the original and transformed frame \cite{stano2012:PRB}).
	
	To calculate the matrix elements of the effective form of the spin-orbit interaction, Eq.~\eqref{eqS:effective soi}, we will use the inversion symmetries of the functions in Eq.~\eqref{eqS:ono 1e basis}. Namely, in the coordinate system where the $x^\prime$ axis is aligned with ${\bf d}$, and $y^\prime$ is perpendicular to it, these functions transform 
	\begin{subequations}
		\begin{eqnarray}
			I_{x^\prime} \Psi_L({\bf r}) = \Psi_R^* ({\bf r}),\qquad I_{x^\prime} \Psi_R({\bf r}) = \Psi_L^* ({\bf r}), \\
			I_{y^\prime} \Psi_L({\bf r}) = \Psi_L^* ({\bf r}),\qquad I_{y^\prime} \Psi_R({\bf r}) = \Psi_R^* ({\bf r}),
		\end{eqnarray}
		\label{eqS:inversion symmetries}
	\end{subequations}
	where $I_{x^\prime}$ is an inversion of the coordinates $(x^\prime, y^\prime) \to (-x^\prime, y^\prime)$ and analogously for $y^\prime$, $^*$ is a complex conjugation, and the dot coordinate system is defined by the angle $\delta$ of the vector ${\bf d}$ and the crystallographic $x$ axis by
	\begin{subequations}
		\begin{eqnarray}
			{x^\prime}  = (x \cos \delta  + y \sin \delta)/\sqrt{2},\\
			{y^\prime}  = (y \cos \delta  - x \sin \delta)/\sqrt{2}.
		\end{eqnarray}
		\label{eqS:dot coordinates}
	\end{subequations}
	Using Eq.~\eqref{eqS:inversion symmetries} it is simple to show that the only non-zero dipole matrix elements within the single electron basis functions we consider are
	\begin{equation}
		\langle R | x^\prime | R\rangle = d = -  \langle L | x^\prime | L\rangle,
		\label{eqS:dipole 1e}
	\end{equation}
	so that the orbital effects of the magnetic field are absent, apart from a small renormalization of the inter-dot overlap $\Omega$, which cancels in $\eta$ anyway.
	Using this result gives that the only non-zero matrix element of the effective spin-orbit interaction, Eq.~\eqref{eqS:effective soi}, within the subspace of states in Eq.~\eqref{eqS:2e basis} is (up to an overall phase coming from the phase convention of the spinors $|\uparrow\rangle$, $|\downarrow\rangle$)
	\begin{equation}
		\langle T_\pm(11) | H_{so}^{eff} | S(11) \rangle = \pm \sqrt{2}d (g \mu_B {\bf B} \times \partial_{x^\prime}{\bf n}_{so})_z.
		\label{eqS:soime}
	\end{equation}
	This gives Eq.~(2) upon identifying
	\begin{equation}
		\lambda_{so}^{-1} = \frac{\cos(\delta-\phi)}{\lambda_B} + \frac{\sin(\delta+\phi)}{\lambda_D},
	\end{equation}
	where we used Eqs.~\eqref{eqS:nso}, \eqref{eqS:dot coordinates} and denoted the angle of the magnetic field with the $x$ axis as $\phi$. This completes the derivation of Eq.~(2).
	
	\section{V. Derivation of Eq.~(5)}
	
	As described in the main text, since the form of the superoperator $L$ in Eq.~(3) is invariant to basis changes within the (11) sector, we choose a basis that diagonalizes the Hamiltonian in this sector. This diagonalization results in a set of states $X$, that are defined by $P(11)HP(11) X = E_X P(11) X$, with $P(11)$ the projector on the (11) subspace. It also results in off-diagonal elements given in Eq.~(4), which we treat perturbatively. Assuming that initially the system is in such a state $X$, we can neglect all other $X^\prime$ states in the lowest order of the time dependent perturbation theory, since the system cannot make a direct tunneling $X\to X^\prime$, as there is no matrix element connecting these states. The only possible tunneling is to $S(02)$, the rate of which is described by a 2 by 2 time dependent Hamiltonian
	\begin{equation}
		H(t)=\left(
		\begin{tabular}{cc}
			$E_X(t)$ & $H_{XS}(t)$\\
			$H_{XS}^\dagger(t) $& $E_S(t)$
		\end{tabular} \right).\\
		\label{eqS:H2}
	\end{equation}
	In this notation, we define 
	\begin{equation}
		\hbar \omega_{XS}=\overline{E_X(t)-E_S(t)},
	\end{equation}
	as the time average of the energy difference, and 
	\begin{equation}
		\Omega(t)= \frac{E_X(t)-E_S(t)}{\hbar} - \omega_{XS},
	\end{equation}
	its temporal fluctuation, with spectral density defined analogously to Eq.~(6),
	\begin{equation}
		\Omega^2(\omega) = \frac{1}{2\pi} \int {\rm d}t \, \overline{\Omega(t)\Omega(0)} \exp( i \omega t).
	\end{equation}
	Assuming that the system is initially in state $X$, and treating the off-diagonal terms in Eq.~\eqref{eqS:H2} as a perturbation, in the leading order of a time dependent perturbation theory we get
	\begin{equation}
		P_S(t) \approx \frac{2\Gamma t}{\hbar^2} e^{-\Omega^2_{p.v.}} \int {\rm d}\omega \, \frac{H_{XS}^2(\omega)}{\Gamma^2+(\omega-\omega_{XS})^2},
		\label{eqS:PS}
	\end{equation}
	where we assumed $\Gamma t \gg 1$ (that is, neglecting the initial oscillations which do not grow with time for long times), and 
	\begin{equation}
		\Omega^2_{p.v.} = \int {\rm d}\omega \, \partial_\omega \Omega^2(\omega)\frac{1-\cos\omega t}{ \omega} \simeq {\rm P} \int {\rm d}\omega \, \frac{\partial_\omega \Omega^2(\omega)}{\omega},
	\end{equation}
	with $P$ standing for principal value of the integral, and the approximation valid if the function $\Omega^2(\omega)$ is smooth on the scale $1/t$. Similarly
	\begin{equation}
		\Gamma = \int {\rm d}\omega \, \Omega^2(\omega)\frac{\sin\omega t}{\omega} \simeq \pi \, \Omega^2(0).
		\label{eqS:Gamma}
	\end{equation}
	If we assume that the principal value term is negligible, the transition rate from state $X$ to $S$ given in Eq.~(5), follows as $\partial_t P_S(t)$ from Eq.~\eqref{eqS:PS}, with the charge decoherence rate given as the spectral density of energy fluctuations at zero frequency, Eq.~\eqref{eqS:Gamma}.
	
	\section{VI. Alternative explanations}
	
	We also considered alternative explanations for the behavior of the observed rates. First of all, one might imagine that once the blocked tunneling becomes too slow, some other mechanism, unrelated to hyperfine or spin-orbit interactions and thus to the magnetic field, takes over in the electron spin relaxation. However, cotunneling, as the only realistic candidate for such a process we are aware of, is expected to result in rates well below the dot-lead rate of 1 Hz which we measured in our dot. Further, the data of Fig.~\ref{Fig_S2}(b) show that the value of the blocked tunneling time is not correlated at all with the appearance and disappearance of the plateau in the ratio of the rates.
	
	Second, we can extend our model to incorporate inelastic tunnelings by adding a noise term, e.g., as a time dependent $e {\bf E}({\bf r},t) \cdot {\bf r}$, into Eq.~ (1). It results in a contribution to the last term in Eq.~(5), through the spectral density $E_{XS}^2(\omega)$, which is in general of a different functional form than the considered $H_{XS}^2(\omega)\propto \delta(\omega)$. Before doing any quantitative estimates, we note that to explain the plateau in the ratio of the rates, the spectral density $E_{XS}^2(\omega)$ would have to be quadratic (assuming the decoherence is low enough such that the Lorentzian denominators cancel). We do not deem this to be a probable explanation, since neither of the electric noise sources which are considered relevant for spin relaxation has such spectral density. Namely, the spectral density of phonons is cubic (piezoelectric) or $\omega^5$ (deformation potential), the 1/f noise is inversely proportional to $\omega$ and the shot and thermal noise are frequency independent. In addition, the dipolar moment of the fluctuating electric field does not couple S(11) and S(02). It only contributes to the decoherence, which our model already takes into account. We expect the sub-dominant (beyond the dipole) moments of the noise fields to be very small. 
	
	Third, we examined the possibility that such additional noise induces inelastic transitions within the (11) sector. A blocked triplet $T_\pm(11)$ in such a scenario first decays to the S(11) and only then tunnels to S(02), instead of making a spin-flip-assisted tunneling directly to S(02). However, as derived in detail in the following subsection, we find that this leads to a decay described by $p_X(t) \approx \exp(-t^2/2T_d T_1)$, which we could clearly distinguish from an exponential decay $p_X(t)=\exp(-t/T_b)$ which we observe for blocked tunnelings. 
	
	\subsection{Inelastic transitions within the (11) sector}
	
	\begin{figure}
		\includegraphics[width=0.45\textwidth]{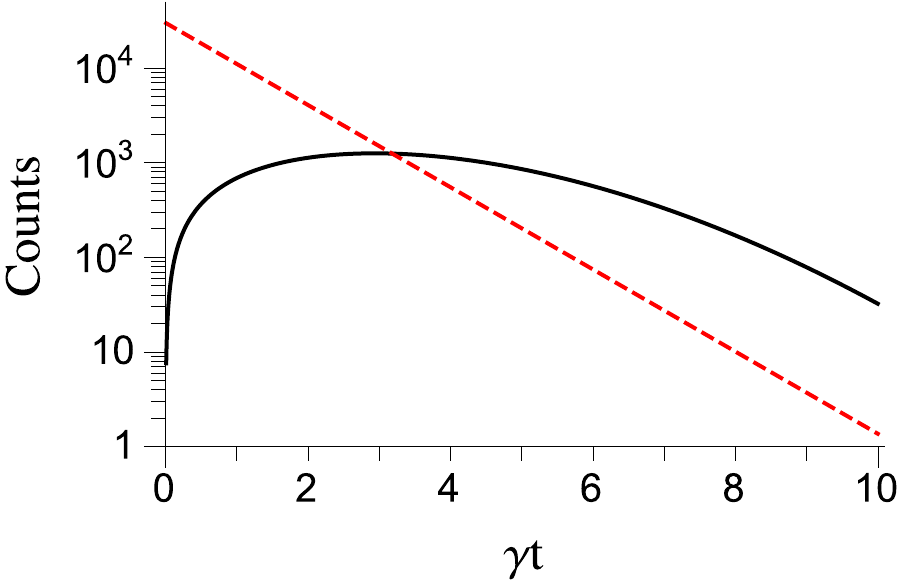}
		\caption{(color online). Histogram simulation. Black solid (red dashed) line gives the contribution from the blocked (direct) tunneling using Eq.~\eqref{eqS:Nd} [Eq.~\eqref{eqS:Nb}]. We assumed that $\langle T_d \rangle = \langle T_b \rangle/20$, and the time average probability for the system to be in a blocked and non-blocked configurations is the same. 
			\label{Fig_S3}}
	\end{figure}
	
	To examine the effects of inelastic transitions, we consider a rate equations model for three states, a polarized triplet denoted as $X$, and the two singlets, $S(11)$ and $S(02)$. The corresponding occupation probabilities are denoted as $p_X$, $p_{11}$, and $p_{02}$, respectively.  They evolve according to
	\begin{subequations}
		\begin{eqnarray}
			\partial_t p_X &=& -W p_X,\\
			\partial_t p_{11} &=& W p_X + \gamma(p_{02}-p_{11}),\\
			\partial_t p_{02} &=& \gamma(p_{11}-p_{02}).
		\end{eqnarray}
		\label{eqS:differential eqs}
	\end{subequations}
	Our minimal model therefore includes a decay of the triplet state into $S(11)$ with relaxation rate $W=1/T_1$ and an unbiased equilibration of the two singlets with the rate $\gamma$, corresponding to the direct inter-dot tunneling $\gamma=1/T_d$. Let us assume that at time $t$ the $S(02)$ was not occupied, $p_{02}(t)=0$. For time interval $\delta t$ short enough, $\delta t \,W, \delta t \,\gamma \ll 1$, Eqs.~\eqref{eqS:differential eqs} are solved by
	\begin{subequations}
		\begin{eqnarray}
			p_X(t+\delta t)& = &p_X(t) \exp(- W \delta t ),\\
			p_{02}(t+\delta t) &=& \gamma \delta t [1-p_X(t)],
		\end{eqnarray}
		\label{eqS:differential eqs solution}
	\end{subequations}
	and the normalization $p_{11}= 1- p_{02}-p_X$, valid at all times exactly.
	
	Now we supplement this evolution by an assumption that the detector measures the system repeatedly with a rate (bandwidth) $1/\delta t$, projecting it between $(02)$ and $(11)$ subspaces. Assuming the system started in (11) subspace, the probability that no tunneling is observed up to time $t$, $P_{11}(t)$, is governed by the equation 
	\begin{equation}
		\delta P_{11} (t + \delta t) = - P_{11}(t) \gamma [1-p_X(t)]  \delta t,
	\end{equation}
	which can be solved, assuming again $\delta t \, W, \delta t \, \gamma \ll 1$, and the initial condition being the state $X$ occupied, $p_X(0)=1$, to give
	\begin{equation}
		P_{11} (t) = \exp \left( -\gamma\frac{Wt - (1-e^{-Wt})}{W}\right).
	\end{equation}
	The condition to observe the system in a blocked configuration is $W\ll\gamma$, otherwise the ``blocked" triplet would decay to $S(11)$ before it could be observed as being blocked. For times $W \ll 1/t \ll \gamma$ we then get from the previous equations the result stated in the main text
	\begin{equation}
		P_{11} (t) \approx \exp \left( -W\gamma t^2/2\right).
	\end{equation}
	The histograms of (11) residing times on Fig.~3 correspond to the probability that the tunneling happens within some interval. According to the previous analysis the number of observed tunnelings within $(t, t+\delta t$) is given by
	\begin{equation}
		N_{b}(t, t+\delta t) = N_0 P_{11}(t) \gamma  \big( 1-\exp(-Wt) \big) \delta t.
		\label{eqS:Nb}
	\end{equation}
	The normalization constant $N_0=T_b^{\rm tot}/\langle T_b \rangle$ is the number of blocked tunnelings, equal to the total time the system spends in a blocked (11) state $T_b^{\rm tot}$ divided by
	\begin{equation}
		\langle T_b \rangle = \int {\rm d} t \, t \partial_t P_{11}(t) \approx \sqrt{\frac{\pi}{2\gamma W}},
	\end{equation}
	an average time to tunnel out from such a state.

	This is to be compared with the initial state of the system in an unblocked (11) state. This situation is described by Eqs.~\eqref{eqS:differential eqs} with the state $X$ never occupied and thus disregarded from considerations. An analogous calculation gives
	\begin{equation}
		N_{d}(t, t+\delta t) = \frac{T_d^{\rm tot}}{\langle T_d \rangle} \gamma \exp \left( -\gamma t \right)  \delta t,
		\label{eqS:Nd}
	\end{equation}
	with $\langle T_d \rangle=1/\gamma$. Assuming the system spends an equal amount of time in a blocked as in an unblocked state, $T_d^{\rm tot}=T_b^{\rm tot}$, we can simulate a histogram plot assuming, for illustration, $\langle T_d \rangle = \langle T_b \rangle/20$, a typical ratio we observe on the plateau. We get the data plotted on Fig.~\ref{Fig_S3}. Comparing with Fig.~3, it demonstrates that we can exclude the possibility of inelastic transitions being relevant in our experiment.

\end{document}